\documentclass[12pt]{article}
\def\timesbox{\hbox{$\scriptscriptstyle\times$}}
\def\ant{ {{\lower 1ex  \timesbox} \atop {\raise 1.5ex  \timesbox}}}
\newcommand{\Zop}{{\hbox{ Z\kern-1.6mm Z}}}
\newcommand{\beq}{\begin{equation}}
\newcommand{\eeq}{\end{equation}}
\newcommand{\bea}{\begin{eqnarray}}
\newcommand{\eea}{\end{eqnarray}}

\newcommand{\lt}{\left}
\newcommand{\rt}{\right}
\newcommand{\Iop}{\relax{\rm I\kern-.18em I}}
\newcommand{\one}{{\hbox{ 1\kern-1.2mm l}}}
\newcommand{\T}{{\cal T}}

\newcommand{\Tt}{{\tilde \T}}

\newcommand{\ltil}{\tilde l}

\newcommand{\Wtil}{\tilde W}
\newcommand{\wtil}{\tilde w}

\newcommand{\dt}{\delta}
\newcommand{\del}{\partial}

\newcommand{\A}{{\cal A}}

\newcommand{\D}{\Delta}
\newcommand{\eps}{\epsilon}

\newcommand{\s}{\sigma}

\begin{document}

{}~
{}~
\hfill\vbox{\hbox{IMSc/2009/07/10}}
\break

\vskip 2cm

\centerline{\Large \bf Dynamical supersymmetry analysis of conformal}
\centerline{\Large \bf invariance for superstrings in type IIB}
\centerline{\Large \bf R-R plane-wave}

\medskip

\vspace*{4.0ex}

\centerline{\large \rm Partha Mukhopadhyay }

\vspace*{4.0ex}

\centerline{\large \it Institute of Mathematical Sciences}
\centerline{\large \it C.I.T. Campus, Taramani}
\centerline{\large \it Chennai 600113, India}

\medskip

\centerline{E-mail: parthamu@imsc.res.in}

\vspace*{5.0ex}

\centerline{\bf Abstract}
\bigskip

In a previous work (arXiv:0902.3750 [hep-th]) we studied the world-sheet conformal invariance for superstrings in type IIB R-R plane-wave in semi-light-cone gauge. Here we give further justification to the results found in that work through alternative arguments using dynamical supersymmetries. We show that by using the susy algebra the same quantum definition of the energy-momentum (EM) tensor can be derived. Furthermore, using certain Jacobi identities we indirectly compute the Virasoro anomaly terms by calculating second order susy variation of the EM tensor. Certain integrated form of all such terms are shown to vanish. In order to deal with various divergences that appear in such computations we take a point-split definition of the same EM tensor. The final results are shown not to suffer from the ordering ambiguity as noticed in the previous work provided the coincidence limit is taken before sending the regularization parameter to zero at the end of the computation.

\newpage

\tableofcontents

\baselineskip=18pt

\section{Introduction}
\label{s:intro}

In a previous work \cite{mukhopadhyay09} we studied world-sheet conformal invariance of type IIB superstrings in R-R plane-wave background \cite{blau01, metsaev01, metsaev02} in semi-light-cone gauge \cite{berkovits04}\footnote{See \cite{slc, slc-anomaly} for earlier work on this gauge.} of Green-Schwarz superstrings\footnote{See \cite{CFT-RR}, \cite{walton} for other approaches for studying world-sheet conformal invariance in this background. World-sheet theories 
for more general pp-waves with R-R flux have been discussed in \cite{RR-pp}.}. To do that we used, following the work of \cite{kazama}, a phase-space operator method where one first defines the exact quantum energy-momentum (EM) tensor on the world-sheet and then explicitly calculates the Virasoro algebra using the basic equal time commutators and  anti-commutators.  

It was argued in \cite{mukhopadhyay09, mukhopadhyay08} that the relevant vacuum of the theory should behave precisely in the same way as the one corresponding to flat background for the operators inside the universal sector as defined in \cite{universality}. It was shown in \cite{mukhopadhyay09} that this is indeed true; however, in order to reproduce the correct physical
spectrum the quantum EM tensor needs to be defined in the following
way: the ``free part'' is ordered according to massless normal
ordering (MNO) (which is the right ordering to be used in flat
background), but the ``interaction part'' according to phase-space
normal ordering (PNO)\footnote{According to PNO the world-sheet fields
  and the conjugate momenta are ordered freely among themselves, but
  have non-trivial mutual ordering \cite{kazama,
    mukhopadhyay09}. However, because of the simplicity of the
  background the EM tensor is independent of this mutual
  ordering. \label{pno}}. Such a definition can be understood to
relate the world-sheet couplings and the space-time fields in a
particular way. To establish conformal invariance the Virasoro algebra
was constructed by directly calculating the commutators of EM tensor
components in local form. It was shown that the anomaly terms  
are well-defined at a finite value of the regularization parameter
$\eps$ and develop an ambiguity in the limit $\eps \to 0$. It is
precisely when the terms are ordered according to PNO at a finite
$\eps$ that the bosonic and fermionic contributions to the anomaly
cancel each other.  

For the above computation it was possible to keep the background metric off-shell in a restricted manner. The bosonic and fermionic contributions mentioned above form the metric and the R-R flux part of the supergravity equation of motion for this restricted ansatz respectively. As mentioned in footnote \ref{pno}, the EM tensor itself is independent of the mutual ordering between the world-sheet fields and the conjugate momenta. However, organizing 
the anomaly terms according to PNO does order such variables in a
particular way. It is not clear to us what this procedure may mean in a more generic context. It is therefore interesting to ask if there exists a generalization of this complete procedure for arbitrary backgrounds. 

In this work we will further justify the results of \cite{mukhopadhyay09} by an analysis using dynamical supersymmetries. The relevant ``currents'' that we work with are related to the usual ones in light-cone gauge \cite{metsaev01,  metsaev02} by a certain scaling. We show that the ``transverse part'' of the EM tensor components, as defined in \cite{mukhopadhyay09}, naturally emerges from the anti-commutators of
such currents. Moreover, we use this result into a number of Jacobi identities involving susy charges and the EM tensor components to relate certain integrated form of the Virasoro anomaly terms to second order susy variations of the EM tensor components. We show that all such total anomaly terms\footnote{Notice that using this procedure, which is valid only onshell, we can not identify the bosonic and ferminic contributions to the anomaly
separately.} vanish without directly encountering any operator ordering ambiguity that is found in the direct method. However, the second order susy variation of the EM tensor components generically gives rise to terms where non-commuting fields appear at the same point. Therefore a reordering performed in such terms, which we need to do to write the final result in the desired form, leads to divergent c-number contributions. In order to deal with such contributions in a systematic manner and to show that the divergences cancel we first perform the computation by point-splitting the EM tensor and then take the coincidence limit at the end. The process of taking such a coincidence limit could potentially encounter operator ordering ambiguity. However, we show that the desired results are obtained unambiguously provided the coincidence limit is imposed before sending the regularization parameter to zero.

The argument using dynamical susy to justify the definition of EM tensor has been discussed in subsection \ref{ss:def} and the computation of Virasoro anomaly has been discussed in subsection \ref{ss:Vir}. The computation using the point-split EM tensor has been discussed in section \ref{s:pt-split}. Various technical details are given in several appendices. 

\section{Dynamical supersymmetry analysis}
\label{susy}

Throughout the paper we will follow the same notations as in \cite{mukhopadhyay09}. We define the right and left moving dynamical supersymmetry currents as, 
\bea
q_{\dot a} = \s^I_{a \dot a} \Pi^I S^a - \mu (\bar \s^I \Sigma)_{\dot
  a a} \chi X^I \tilde  S^a~,  \quad 
\tilde q_{\dot a} = \s^I_{a \dot a} \tilde \Pi^I \tilde S^a + \mu 
(\bar \s^I \Sigma)_{\dot a a} \chi X^I S^a~,
\label{q-qtilde-def}
\eea
respectively, where $\chi=\sqrt{\Pi^+\tilde \Pi^+}$. Notice that there is no normal ordering ambiguity in the above definitions. Below we use the corresponding charges\footnote{We use the notation $\oint d\s$ to denote the definite integral $\int_0^{2\pi}d\s $. An indefinite integral will be denoted $\int d\s$, as usual.}
$Q_{\dot a}= \oint {d\s \over 2\pi} ~q_{\dot a}(\s)$ and $\tilde Q_{\dot
  a}= \oint {d\s \over 2\pi} ~\tilde q_{\dot a}(\s)$ to justify the
definition of the EM tensor given in \cite{mukhopadhyay09} (and as
summarized in appendix \ref{a:def}) and that the theory is free from
conformal anomaly. 

\subsection{Definition of EM tensor}
\label{ss:def}

Given the currents in eqs.(\ref{q-qtilde-def}) the algebra of charges
turn out to be: 
\bea
 \{Q_{\dot a}, Q_{\dot b} \} = 2 \delta_{\dot a \dot b}  \oint {d\s
   \over 2 \pi} ~\T_{\perp}(\s)~,   \quad
\{\tilde Q_{\dot a}, \tilde
 Q_{\dot b} \} = 2 \delta_{\dot a \dot b}  \oint {d\s \over 2 \pi}
 ~\tilde \T_{\perp}(\s)~,
\label{susy-alg}
\eea
where the transverse components of the EM tensor, namely $\T_{\perp}$ and $\Tt_{\perp}$ have been defined in appendix \ref{a:def}. The above results can be easily derived as follows. Considering the right-moving sector first, a straightforward computation shows\footnote{Throughout the paper we will use the notation: $\D = \s-\s'$.}: 
\bea
\{q_{\dot a} (\s), q_{\dot b}(\s')\}
&=& \{q^{(0)}_{\dot a} (\s), q^{(0)}_{\dot b}(\s')\} + 4\pi \delta_{\dot a \dot b} \delta \T (\s) \delta_{\eps}(\Delta)~, 
\label{q-q-anti-comm}
\eea
where we have used: $\s^I_{a\dot a}\s^I_{b\dot b}+ \s^I_{a\dot b}\s^I_{b\dot a}= 2\delta_{ab} \delta_{\dot a \dot b} $, $\delta \T$ has been defined in appendix \ref{a:def} and $q^{(0)}_{\dot a}(\s)$ is the free part obtained by setting $\mu=0$ in eq(\ref{q-qtilde-def}). We will not attempt to compute the local anti-commutator $\{q^{(0)}_{\dot a} (\s), q^{(0)}_{\dot b}(\s')\}$ here, rather we will use the standard result for the corresponding charges \cite{gsw}:
\bea
\{Q^{(0)}_{\dot a}, Q^{(0)}_{\dot b}\} = 2\delta_{\dot a \dot b} \oint {d\s \over 2\pi}~ \T_{\perp}^{(0)}(\s)~,
\label{susy-alg-0}
\eea
where $\T_{\perp}^{(0)}$ is the free part of $\T_{\perp}$ (see appendix \ref{a:def}). The first equation of (\ref{susy-alg}) then directly follows from eqs.(\ref{q-q-anti-comm}) and (\ref{susy-alg-0}). The argument for the left-moving sector is similar. 

The results in (\ref{susy-alg}) give an alternative justification that the interaction term $\dt \T$ in the EM tensor be defined according to PNO. The argument goes as follows. We may express the right hand side of the first equation in (\ref{susy-alg}) in the following way: 
\bea
\{Q_{\dot a}, Q_{\dot b}\} = 2 \delta_{\dot a \dot b} \oint {d\s \over
  2\pi} \lt[ \T_{\perp}^{MNO}(\s) + {d \mu^2\over 2} D_{\eps}(0)
\chi^2(\s) \rt]~,  
\label{susy-MNO}
\eea
(and similarly for the left moving sector) where the number of transverse directions is $d=8$. $D_{\eps}(\D)$ is defined through the following equation: 
\bea
\ant X^{\mu}(\s) X^{\nu}(\s')\ant &=& :X^{\mu}(\s) X^{\nu}(\s'): +\eta^{\mu \nu} D_{\eps}(\D)~, \label{propagator-def}
\eea
where $\ant~\ant$ and $:~:$ denote PNO and MNO respectively and it evaluates to be \cite{mukhopadhyay09}, 
\bea
D_{\eps}(\D) &=& {i\over 4\pi T} \int d\D
\lt(d(e^{i\D},\eps)-d(e^{-i\D},\eps) \rt) ~, \cr
&=& -{1\over 4\pi T}\ln (\D^2+\eps^2)~. \label{D}
\eea
$\T_{\perp}^{MNO}(\s)$ is same as $\T_{\perp}(\s)$ with the interaction term defined according to MNO. It turns out that the second term on the right hand side of eq.(\ref{susy-MNO}) evaluates, inside the transverse Hilbert space ${\cal H}_{\perp}$ (as defined in \cite{mukhopadhyay09}), to be positively divergent: ${d(\alpha'\mu p^+)^2 \over 4} \sum_{n>0} {e^{-n\eps}\over n}$, where we have used the first line in eq.(\ref{D}). Therefore the supersymmetry algebra in eqs.(\ref{susy-alg}) implies that it is the EM tensor defined
following the PNO prescription that leads to a positive definite spectrum. 

\subsection{Computation of Virasoro anomaly}
\label{ss:Vir}

Here we would like to compute certain integrated form of the Virasoro anomaly terms as defined in \cite{mukhopadhyay09} indirectly by using the following Jacobi identities:
\bea
\lt\{[\T_{\perp}(\s), Q_{\dot a}], Q_{\dot b} \rt\} +
\lt\{[\T_{\perp}(\s), Q_{\dot b}], Q_{\dot a} \rt\} &=&
\lt[\T_{\perp}(\s), \{Q_{\dot a}, Q_{\dot b} \}\rt] \cr 
&=& 2\delta_{\dot a \dot b} \oint {d\s' \over 2\pi} [\T_{\perp}(\s),
\T_{\perp}(\s')]~, \cr &&  
\label{jacobiR} \\
\lt\{[\tilde \T_{\perp}(\s), \tilde Q_{\dot a}], \tilde Q_{\dot b} \rt\} +
\lt\{[\tilde \T_{\perp}(\s), \tilde Q_{\dot b}], \tilde Q_{\dot a}
\rt\} &=& \lt[\tilde \T_{\perp}(\s), \{\tilde Q_{\dot a}, \tilde
Q_{\dot b} \}\rt] \cr 
&=& 2\delta_{\dot a \dot b} \oint {d\s' \over 2\pi} [\tilde
\T_{\perp}(\s), \tilde \T_{\perp}(\s')]~, \cr &&  
\label{jacobiL} \\
\lt\{[\tilde \T_{\perp}(\s), Q_{\dot a}], Q_{\dot b} \rt\} +
\lt\{[\tilde \T_{\perp}(\s), Q_{\dot b}], Q_{\dot a} \rt\} &=&
\lt[\tilde \T_{\perp}(\s), \{Q_{\dot a}, Q_{\dot b} \}\rt] \cr 
&=& 2\delta_{\dot a \dot b} \oint {d\s' \over 2\pi} [\tilde
\T_{\perp}(\s), \T_{\perp}(\s')]~, 
\label{jacobiRL}
\eea
where in the second line of each of eqs.(\ref{jacobiR}, \ref{jacobiL}, \ref{jacobiRL}) we have used the susy algebra (\ref{susy-alg}). We will compute the two sides of the above equations independently. The right hand sides, which involve the ``transverse parts'' of the Virasoro algebra, will contain the anomaly terms. The relevant expressions are given by (see appendix \ref{a:trans} for derivation): 
\bea
\lt[\T_{\perp}(\s), \T_{\perp}(\s') \rt] &=& {d \pi i \over 4}
\delta'''(\Delta) - 4\pi i \lt(\T^{(0)}_{\perp}(\s) + {1\over 2}
\delta \T_F(\s)\rt) \delta'(\Delta) \cr 
&& -2\pi i \del  \lt(\T^{(0)}_{\perp}(\s) + {1\over 2} \delta
\T_F(\s) \rt) \delta (\Delta) + {\cal A}^R(\s,\s')~,  \cr &&   
\label{Tperp-TperpR} \\
\lt[\tilde \T_{\perp}(\s), \tilde \T_{\perp}(\s') \rt] &=& - {d \pi i \over
  4} \delta'''(\Delta) + 4\pi i \lt(\tilde \T^{(0)}_{\perp}(\s) +
{1\over 2} \delta \T_F(\s) \rt) \delta'(\Delta) \cr  && 
+2\pi i \del  \lt(\tilde \T^{(0)}_{\perp}(\s) + {1\over 2}
\delta \T_F(\s) \rt) \delta (\Delta) + {\cal A}^L(\s,\s')~,  \cr &&  
\label{Tperp-TperpL} \\
\lt[\T_{\perp}(\s), \tilde \T_{\perp}(\s') \rt] &=& {\mu \over 2} \sqrt{\pi
  \over T} \lt[\del  \chi (\s) (S\Sigma \tilde S(\s)) - \chi(\s)
\del (S\Sigma \tilde S(\s))\rt] \delta(\Delta) \cr && 
- \pi i \chi^2(\s) \del  K(\vec X(\s)) \delta (\Delta)+ {\cal A}(\s,\s') ~.
\label{Tperp-TperpRL}
\eea
where the anomaly terms ${\cal A}^R(\s,\s')$, ${\cal A}^L(\s,\s')$ and ${\cal A}(\s,\s')$ are as defined in appendix \ref{a:trans}. By integrating the above results and using them in eqs.(\ref{jacobiR}), (\ref{jacobiL}) and (\ref{jacobiRL}) respectively we arrive at: 
\bea
\lt\{[\T_{\perp}(\s), Q_{\dot a}], Q_{\dot b} \rt\} +
\lt\{[\T_{\perp}(\s), Q_{\dot b}], Q_{\dot a} \rt\} &=&
-2i\delta_{\dot a \dot b} \del \lt(\T_{\perp}^{(0)}(\s) +{1\over
  2} \delta \T_F(\s) \rt) \cr 
&& + 2\delta_{\dot a\dot b} \oint {d\s' \over 2\pi} ~{\cal
  A}^R(\s,\s')~, \cr && \label{jacobiR2} \\ 
\lt\{[\tilde \T_{\perp}(\s), \tilde Q_{\dot a}], \tilde Q_{\dot b} \rt\} +
\lt\{[\tilde \T_{\perp}(\s), \tilde Q_{\dot b}], \tilde Q_{\dot a} \rt\} &=&
2i\delta_{\dot a \dot b} \del \lt(\tilde \T_{\perp}^{(0)}(\s)
+{1\over 2} \delta \T_F(\s) \rt) \cr 
&& + 2\delta_{\dot a\dot b} \oint {d\s' \over 2\pi} ~{\cal
  A}^L(\s,\s')~, \cr && \label{jacobiL2} \\ 
\lt\{[\tilde \T_{\perp}(\s), Q_{\dot a}], Q_{\dot b} \rt\} +
\lt\{[\tilde \T_{\perp}(\s), Q_{\dot b}], Q_{\dot a} \rt\} &=&
{\mu \over 2 \sqrt{\pi T}} \delta_{\dot a \dot b} \lt[ \chi(\s) \del (S\Sigma \tilde S(\s))\rt. \cr 
&& \lt. - \del \chi (\s) (S\Sigma \tilde S(\s)) \rt] +i
\delta_{\dot a \dot b} \chi^2(\s) \del K(\vec X(\s)) \cr 
&& - 2\delta_{\dot a\dot b} \oint {d\s' \over 2\pi} {\cal A}(\s',\s)~. \label{jacobiRL2} 
\eea
The idea is to evaluate the integrated forms of the Virasoro anomaly terms that appear on the right hand sides of the above equations by independently computing the second order susy variations of the EM tensor components that appear on the left hand sides. Below we describe how such second order variations are obtained and what the results are.

The first order susy variations of the basic fields and various parts of $\T_{\perp}$ and $\Tt_{\perp}$ have been calculated in appendix \ref{a:1st}. Using these results one finds:
\bea
\lt[\T_{\perp}(\s), Q_{\dot a}\rt] &=&-{i\over 2}\s^I_{a\dot a} \del \lt(\Pi^I(\s) S^a(\s)\rt)~, \label{1stR} 
\\ 
\lt[\tilde \T_{\perp}(\s), \tilde Q_{\dot a}\rt] &=&{i\over 2}\s^I_{a\dot a} \del \lt(\tilde \Pi^I(\s) \tilde S^a(\s)\rt)~, \label{1stL} 
\\ 
\lt[\tilde \T_{\perp}(\s), Q_{\dot a}\rt] &=& i\mu (\bar \s^I\Sigma)_{\dot a a} \chi(\s) \del X^I(\s) \tilde S^a(\s)\cr
&&  +{i\mu \over 2} (\bar \s^I\Sigma)_{\dot a a}\lt[\chi(\s)X^I(\s)\del \tilde S^a(\s) - \del \lt(\chi(\s)X^I(\s) \rt) \tilde S^a(\s) \rt] ~. \label{1stRL}
\eea
Given the above expressions the second order susy variations can be computed using the results (\ref{basic-susy-R}) and (\ref{basic-susy-L}). As mentioned earlier, there is a subtlety in this derivation which we will explain toward the end of this section. The final results are,
\bea
\{[\T_{\perp}(\s), Q_{\dot a}], Q_{\dot b}\} + \{[\T_{\perp}(\s),
Q_{\dot b}], Q_{\dot a}\} &=& -2i \delta_{\dot a \dot b}\del
\lt(\T_{\perp}^{(0)}(\s)+{1\over 2}\delta \T_F(\s) \rt)~, \label{lhsR}
\\ 
\{[\tilde \T_{\perp}(\s), \tilde Q_{\dot a}], \tilde Q_{\dot b}\} +
\{[\tilde \T_{\perp}(\s), \tilde Q_{\dot b}], \tilde Q_{\dot a}\} &=&
2i \delta_{\dot a \dot b}\del \lt(\tilde
\T_{\perp}^{(0)}(\s)+{1\over 2}\delta \T_F(\s) \rt)~, \label{lhsL} \\ 
\{[\tilde \T_{\perp}(\s), Q_{\dot a}], Q_{\dot b}\} + \{[\tilde
\T_{\perp}(\s), Q_{\dot b}], Q_{\dot a}\}&=& 
{\mu \over 2\sqrt{\pi T}} \delta_{\dot a \dot b} \lt[\chi(\s) \del (S\Sigma \tilde S(\s)) \rt.  \cr
&& \lt. - \del \chi(\s) (S\Sigma \tilde S(\s)) \rt] + i \delta_{\dot a \dot b} \chi^2(\s) \del K(\vec
X(\s))~.  \cr && \label{lhsRL} 
\eea
Substituting these results into eqs.(\ref{jacobiR2}, \ref{jacobiL2}, \ref{jacobiRL2}) we conclude,  
\bea
\oint d\s' \A^R(\s,\s') = \oint d\s' \A^L(\s,\s')= \oint d\s' \A(\s',\s) = 0~.
\eea

We will now discuss the subtlety involved in deriving equations (\ref{lhsR}), (\ref{lhsL}) and (\ref{lhsRL}) from equations (\ref{1stR}), (\ref{1stL}) and (\ref{1stRL}) respectively. Given the susy variations of the basic fields in (\ref{basic-susy-R}) and (\ref{basic-susy-L}), it is clear that a further susy transformation of (\ref{1stR}) and (\ref{1stL}) will give rise to terms where non-commuting local fields appear at the same point. In order to give the results the forms of the right hand sides of eqs.(\ref{lhsR}, \ref{lhsL}) one needs to reorder such terms in a particular way. Such a procedure leads to divergent c-number contributions as the fields are coincident. In order to do the computation more systematically we will point-split the EM tensor and take the coincidence limit at the end of the computation. This analysis will be discussed in the next section where we show that if the coincidence limit is performed before taking the regularization parameter $\eps$ to zero, then the results (\ref{lhsR}) and (\ref{lhsL}) follow unambiguously. Notice that the subtlety discussed above does not apply to eq.(\ref{1stRL}). This is because that a further susy transformation leads to terms where only commuting fields are coincident.  Therefore the derivation of (\ref{lhsRL}) from (\ref{1stRL}) is straightforward. However, as a consistency check we will show in the next section that the corresponding point-split analysis leads to the same conclusion only when the coincidence limit is performed the way it has been described above.

\section{Computation with point-split EM tensor}
\label{s:pt-split}

Here we will describe in detail the point-split computation for the derivation of eqs.(\ref{lhsR}, \ref{lhsL}, \ref{lhsRL}) as summarized at the end of the previous section. We define the symmetrized point-split version of the EM tensor as follows:
\bea
\T^{\beta}_{\perp}(\s) &=& t^{\beta}(\s) + s^{\beta}(\s) + \dt \T^{\beta}_B(\s) + \dt \T^{\beta}_F(\s)~, \cr
\Tt^{\beta}_{\perp}(\s) &=& \tilde t^{\beta}(\s) + \tilde s^{\beta}(\s) + \dt \T^{\beta}_B(\s) + \dt \T^{\beta}_F(\s)~,
\label{Tbeta}
\eea
where,
\bea
t^{\beta}(\s) &=& {1\over 4} \lt[\Pi^I(\s+\beta) \Pi^I(\s) + \Pi^I(\s) \Pi^I(\s+\beta)\rt]~, \cr 
s^{\beta}(\s) &=& -{i\over 4} \lt[S^a(\s+\beta) \del S^a(\s) + S^a(\s)\del S^a(\s+\beta)\rt]~, \cr
\tilde t^{\beta}(\s) &=& {1\over 4} \lt[\tilde \Pi^I(\s+\beta) \tilde \Pi^I(\s) + \tilde \Pi^I(\s) \tilde \Pi^I(\s+\beta)\rt]~, \cr 
\tilde s^{\beta}(\s) &=& {i\over 4} \lt[\tilde S^a(\s+\beta) \del \tilde S^a(\s) + \tilde S^a(\s) \del \tilde S^a(\s+\beta)\rt]~,\cr
\dt \T^{\beta}_B(\s) &=& {\mu^2\over 2} \chi^2(\s) X^I(\s+\beta) X^I(\s)~, \cr
\dt \T^{\beta}_F(\s) &=& {i\mu \over 4 \sqrt{\pi T}}\chi(\s) \lt[(S(\s+\beta)\Sigma \tilde S(\s)) + (S(\s)\Sigma \tilde S(\s+\beta)) \rt]~.
\label{Tbeta-parts}
\eea
The actual EM tensor defined in appendix \ref{a:def} is obtained by taking the coincidence limit:
\bea
\T_{\perp}(\s) = \lim_{\beta \to 0} \T^{\beta}_{\perp}(\s)~,\quad 
\Tt_{\perp}(\s) = \lim_{\beta \to 0} \Tt^{\beta}_{\perp}(\s)~.
\eea
Notice that the ordering of operators in $\dt \T^{\beta}_B$ and $\dt \T^{\beta}_F$ are same as in $\dt \T_B$ and $\dt \T_F$ respectively and therefore there is no divergent contribution coming from such terms in the coincidence limit. This, however, is not true for the free parts $t^{\beta}(\s)$, $s^{\beta}(\s)$, $\tilde t^{\beta}(\s)$ and $\tilde s^{\beta}(\s)$ separately. But the divergences cancel between the bosonic and fermionic parts, as expected, because of the following relations:
\bea
\lim_{\beta \to 0} t^{\beta}(\s) &=&  t(\s) + \lim_{\beta \to 0} {d\over 4} \lt[{1\over (\eps +i\beta)^2} + {1\over (\eps -i\beta)^2} \rt]~, \cr
\lim_{\beta \to 0} s^{\beta}(\s) &=& s(\s) - \lim_{\beta \to 0} {d\over 4} \lt[{1\over (\eps +i\beta)^2} + {1\over (\eps -i\beta)^2} \rt]~.
\eea
Similar relations also hold for the left moving sector. This is easily derived from the short distance behavior of the basic fields as described in appendix \ref{a:ope}. We will compute the relevant susy variations for the point-split quantities and then finally take the coincidence limit to arrive at eqs.(\ref{lhsR}), (\ref{lhsL}) and (\ref{lhsRL}). Using the susy variations of the basic fields as given in eqs.(\ref{basic-susy-R}) and (\ref{basic-susy-L}) one derives the following results:
\bea
\lt[\T^{\beta}_{\perp}(\s), Q_{\dot a}\rt] &=& -{i\over 4} \s^I_{a \dot a} \del \lt(\Pi^I(\s) S^a(\s+\beta) + \Pi^I(\s+\beta) S^a(\s)\rt) \cr
&& + {i\mu \over 4\sqrt{\pi T}} (\bar \s^I\Sigma)_{\dot a a} \lt(\chi(\s+\beta) -\chi(\s) \rt) 
\Pi^I(\s) \tilde S^a(\s +\beta) \cr
&& - {i\mu^2 \over 4\sqrt{\pi T}}\s^I_{a\dot a} \lt(\chi(\s+\beta) -\chi(\s)\rt) \chi(\s) X^I(\s+\beta) S^a(\s)  ~,  \label{Tbeta-Q} \\ && \cr
\lt[\tilde \T^{\beta}_{\perp}(\s), \tilde Q_{\dot a}\rt] &=& {i\over 4} \s^I_{a\dot a} \del \lt(\tilde \Pi^I(\s) \tilde S^a(\s+\beta)  + \tilde \Pi^I(\s+\beta) \tilde S^a(\s) \rt) \cr
&& -{i \mu \over 4\sqrt{\pi T}} (\bar \s^I \Sigma)_{\dot a a} \lt(\chi(\s+\beta)-\chi(\s) \rt)\tilde \Pi^I(\s) S^a(\s+\beta) \cr
&& -{i\mu^2 \over 4\sqrt{\pi T}} \s^I_{a\dot a} \lt(\chi(\s+\beta) -\chi(\s)\rt) \chi(\s) X^I(\s+\beta) \tilde S^a(\s)~, \label{Ttbeta-Qt} \\ && \cr
\lt[\tilde \T^{\beta}_{\perp}(\s), Q_{\dot a}\rt] &=& {i\mu \over 4 \sqrt{\pi T}} (\bar \s^I\Sigma)_{\dot a a} \lt[\chi(\s) \tilde \Pi^I(\s+\beta)  \tilde S^a(\s) + \chi(\s+\beta) \tilde \Pi^I(\s) \tilde S^a(\s+\beta) \rt. \cr
&& \lt. -\chi(\s) \Pi^I(\s+\beta) \tilde S^a(\s) - \chi(\s) \Pi^I(\s) \tilde S^a(\s+\beta)\rt] \cr
&& + {i\mu \over 4} (\bar \s^I\Sigma)_{\dot a a} \lt\{ \chi(\s+\beta) X^I(\s+\beta) \del \tilde S^a(\s) +\chi(\s)X^I(\s) \del \tilde S^a(\s+\beta) \rt. \cr
&& \lt. - \del \lt(\chi(\s) X^I(\s) \rt) \tilde S^a(\s+\beta) - \del \lt(\chi(\s+\beta) X^I(\s+\beta) \rt) \tilde S^a(\s)\rt\} \cr
&& -{i\mu^2\over 4\sqrt{\pi T}} \s^I_{a\dot a} \lt(\chi(\s+\beta)-\chi(\s)\rt) \chi(\s) X^I(\s+\beta) S^a(\s) ~.\cr 
&& \label{Ttbeta-Q}
\eea
Below we will discuss the relevant second order variation of equations (\ref{Tbeta-Q}), (\ref{Ttbeta-Qt}) and (\ref{Ttbeta-Q}) in the same order and show how, in the coincidence limit, they reduce to equations (\ref{lhsR}), (\ref{lhsL}) and (\ref{lhsRL}) respectively. 

We will start the discussion with eq.(\ref{Tbeta-Q}). Computing the susy variation of the first term with respect to $Q_{\dot b}$ and symmetrizing in $\dot a$ and $\dot b$ one obtains:
\bea
&&\del \lt[-{i\over 4} \s^I_{a\dot a} \{ \Pi^I(\s)S^a(\s+\beta)+\Pi^I(\s+\beta)S^a(\s), Q_{\dot b} \} + \dot a \leftrightarrow \dot b\rt] \cr &=& -2i\dt_{\dot a \dot b} \del \lt[ t^{\beta}(\s) + s^{\beta}(\s) 
 + {i\mu \over 8\sqrt{\pi T}} \chi(\s) \lt(S(\s+\beta)\Sigma \tilde S(\s) \rt) \rt. \cr 
&& \lt. + {i\mu \over 8\sqrt{\pi T}} \chi(\s+\beta) \lt(S(\s)\Sigma \tilde S(\s+\beta) \rt) \rt] ~, \cr
&\stackrel{\beta \to 0}{\to}& -2i \dt_{\dot a \dot b} \del_{\s} \lt(\T^{(0)}_{\perp} + {1\over 2} \dt \T_F(\s) \rt)~,
\eea
where in the last line the coincidence limit has been taken and the resulting contribution gives the right hand side of eq.(\ref{lhsR}). Therefore we need to show that the susy variation of the rest of the terms in eq.(\ref{Tbeta-Q}) with $Q_{\dot b}$, symmetrized between $\dot a$ and $\dot b$, vanishes in the coincidence limit. Since the factor $\phi^{\beta}(\s) = (\chi(\s+\beta)- \chi(\s))$ does not vary under susy, we need to consider the variations of $A^{Ia}_{\beta}(\s) = \Pi^I(\s) \tilde S^a(\s+\beta)$ and $B^{Ia}_{\beta}(\s)=X^I(\s+\beta)S^a(\s)$ only. Any term in such variations which has coincident commuting fields will go to zero unambiguously in the coincidence limit because of the factor $\phi^{\beta}(\s)$. Terms which have coincident non-commuting fields can lead to non-vanishing results in the coincidence limit depending on how they are ordered. From eqs.(\ref{basic-susy-R}) it is clear that such terms arising from the susy variations of $A^{Ia}_{\beta}(\s)$ and $B^{Ia}_{\beta}(\s)$ are of the forms $\tilde S(\s) \tilde S(\s+\beta)$, $S(\s+\beta) S(\s)$, $\Pi(\s) X(\s+\beta)$ and $X(\s+\beta)\Pi(\s)$ (we dropped space-time indices to show the forms of the terms schematically). Any such term, after being ordered according to MNO or PNO, will always lead to a short distance behavior of the form ${\beta \over \eps \pm i \beta}$, where the factor of $\beta$ in the numerator comes from the factor of $\phi^{\beta}(\s)$. Therefore if we perform the coincidence limit before sending the regularization parameter $\eps$ to $0$ we always get a vanishing result. 

The argument goes in a similar way for the left moving sector, i.e. eq.(\ref{Ttbeta-Qt}). Let us now consider eq.(\ref{Ttbeta-Q}). As mentioned at the beginning of this section the derivation of (\ref{lhsRL}) from (\ref{1stRL}) is straightforward and does not require one to consider the point-split EM tensor.  However, we will now argue that in order to arrive at the result (\ref{lhsRL}) we need to take the coincidence limit before sending $\eps$ to zero. To do that let us first notice that the terms inside the square brackets in the first two lines of eq.(\ref{Ttbeta-Q}) can be written in the following form:
\bea
&& 2\sqrt{\pi T} \chi(\s) \lt[\del X^I(\s+\beta) \tilde S^a(\s) + \del X^I(\s)\tilde S^a(\s+\beta)\rt] + \beta \del \chi(\s) \tilde \Pi^I(\s) \tilde S^a(\s+\beta) + {\cal O}(\beta^2)~, \cr &&
\eea
where we have used: $\tilde \Pi^I-\Pi^I =2\sqrt{\pi T}\del X^I$. Susy variation of the terms inside the square brackets in the above expression and those inside the curly brackets in eq.(\ref{Ttbeta-Q}) gives rise to product of commuting operators. Therefore the coincidence limit can be taken before performing the susy variation. All these terms together constitute the right hand side of (\ref{1stRL}). Susy variation of the ${\cal O}(\beta)$ term in the above expression 
and the last term in eq.(\ref{Ttbeta-Q}) does produce non-commuting operators. But because of the factor of $\beta$ in the first case and the factor of $\phi^{\beta}(\s)$ in the second the contributions will vanish if we take the coincidence limit first as mentioned earlier. This shows that the second order susy variation of eq.(\ref{Ttbeta-Q}) will lead to 
(\ref{lhsRL}) if the coincidence limit is taken the way we have described here. 

\section{Conclusion}
\label{s:conclusion}

In this paper we have given alternative arguments using dynamical supersymmetries to arrive at the same results as found in \cite{mukhopadhyay09}. In particular we have shown that the anti-commutator of the supercharges naturally leads to the way the quantum EM tensor was defined in \cite{mukhopadhyay09}. Using the susy algebra it can be argued that the EM tensor defined this way leads to a positive definite spectrum. We have also evaluated certain integrated forms of the Virasoro anomaly terms defined in \cite{mukhopadhyay09} indirectly using a number Jacobi identities where one computes second order susy variations of the EM tensor components. These computations have been done carefully by point-splitting the EM tensor and all the anomaly terms mentioned above have been shown to vanish. In this method one does not directly encounter the operator ordering ambiguity as found in \cite{mukhopadhyay09} where a direct method of calculating the Virasoro anomaly terms was adopted. 

\begin{center}
{\bf Acknowledgement}
\end{center}

I thank Suresh Govindarajan for useful discussion.

\appendix

\section{Definition of EM tensor}
\label{a:def}

Here we summarize the definition of the EM tensor as given in
\cite{mukhopadhyay09}. This will also serve the purpose of introducing certain new definitions that we will use later in this paper. The right and left moving parts are given by,
\bea
\T = l+\T_{\perp}+1~, \quad \tilde \T = \tilde l + \tilde \T_{\perp}+1~,
\label{T-parts}
\eea
where the longitudinal parts are given by,
\bea
l=W + \xi w~, \quad \ltil = \Wtil + \xi \wtil~,
\label{l-def}
\eea
with $\xi=-{1\over 2}$, and\footnote{Throughout the paper we will use the notation $\del$ to indicate derivative with respect to $\s$.}
\bea
W = :\Pi^+\Pi^-:~, \quad \Wtil = :\tilde \Pi^+\tilde \Pi^-:~, \quad
w=\del^2\ln \Pi^+~, \quad \wtil = \del^2 \ln \tilde \Pi^+~.
\eea
The transverse parts, on the other hand, are given by,
\bea
\T_{\perp}=\T^{(0)}_{\perp}+\dt \T ~, \quad \Tt_{\perp}=\Tt^{(0)}_{\perp}+\dt \T~.
\label{Tperp}
\eea
The transverse components in flat background are given by,
\bea
\T^{(0)}_{\perp}= t + s~, \quad \Tt^{(0)}_{\perp} = \tilde t + \tilde s~,
\label{Tperp-flat}
\eea
where
\bea
t={1\over  2} :\Pi^I\Pi^I:~, \quad s=-{i\over 2} :S\del S:~,  \quad
\tilde t={1\over  2} :\tilde \Pi^I\tilde \Pi^I:~, \quad 
\tilde s={i\over 2} :\tilde S\del \tilde S:~.
\eea
All these operators are defined with MNO. The ``interaction part'', on
the other hand is defined with PNO:
\bea
\delta \T = \delta \T_B +\delta \T_F~, \quad \delta \T_B = -{1\over 2}
\chi^2 K(\vec X)~, \quad \delta \T_F = {i\mu \over 2\sqrt{\pi T}}\chi
(S\Sigma \tilde S) ~, 
\eea
where,
\bea
K(\vec x) = -\mu^2 \vec x^2 ~, \quad \chi = \sqrt{\Pi^+ \tilde \Pi^+}~,
\eea
and $\Sigma = \s^{1234}$ is the product of $SO(8)$ Dirac
matrices along the directions $I=1,2,3,4$.

\section{Transverse parts of Virasoro algebra}
\label{a:trans}

$\T(\s)$ and $\Tt(\s)$ are expected to satisfy right and left moving
Virasoro algebra with central charge $c=26$. The anomaly terms $\A^R(\s,\s')$, $\A^L(\s,\s')$ and $\A(\s,\s')$ are defined by: 
\bea
[\T(\s), \T(\s')] &=& \pi i\lt[{c\over 6}\delta'''(\D)- \lt(4\T(\s)
-{c\over 6} \rt)\delta'(\D) - 2 \del T(\s) \delta(\D) \rt] +
\A^R(\s,\s') ~, \cr 
[\Tt(\s), \Tt(\s')] &=& -\pi i\lt[{c \over 6}\delta'''(\D)
- \lt(4\Tt(\s) -{c\over 6} \rt)\delta'(\D)
- 2 \del \Tt(\s) \delta(\D) \rt] + \A^L(\s,\s')~, \cr
[\T(\s), \Tt(\s')] &=& \A(\s,\s')~,
\label{vir-alg}
\eea
where,
\bea
c= {3d\over 2} +2-24 \xi = 26~. 
\eea
The longitudinal parts $l$ and $\tilde l$ (see eqs.(\ref{T-parts}, \ref{l-def})) satisfy 
Virasoro algebra with central charge $2-24 \xi= 14$ : 
\bea
[l(\s), l(\s')] &=& \pi i\lt[{2-24 \xi \over 6}\delta'''(\D)- \lt(4l(\s)
-{2\over 6} \rt)\delta'(\D) - 2 \del l(\s) \delta(\D) \rt]~, \cr
[\ltil(\s), \ltil(\s')] &=& -\pi i\lt[{2-24 \xi \over 6}\delta'''(\D)
- \lt(4\ltil(\s) -{2\over 6} \rt)\delta'(\D)
- 2 \del \ltil(\s) \delta(\D) \rt]~, \cr
[l(\s), \ltil (\s')] &=& 0~.
\label{l-ltil-comm}
\eea
These results can be obtained by first noticing that $W$ and $\tilde W$ must satisfy the following right and left moving Virasoro algebra respectively with central charge $2$,
\bea
[W(\s), W(\s')] &=& \pi i \lt[{2\over 6} \delta'''(\D) - \lt(4
W(\s)-{2\over 6} \rt) \delta'(\D) - 2\del W(\s) \delta (\D) \rt]~,
\cr
[\tilde W(\s), \tilde W(\s')] &=& -\pi i \lt[{2\over 6} \delta'''(\D)
- \lt(4 \Wtil(\s)-{2\over 6} \rt) \delta'(\D) -
2\del \Wtil(\s) \delta (\D) \rt]~,
\eea
and using the following commutators:
\bea
[W(\s), w(\s')] &=& 2\pi i \lt[-\delta'''(\D) + \del \ln \Pi^+(\s)
\delta''(\D) \rt]~, \cr
[\Wtil(\s), \wtil(\s')] &=& -2\pi i \lt[-\delta'''(\D) + \del
\ln \tilde \Pi^+(\s)\delta''(\D) \rt]~.
\label{Ww-comm}
\eea
Commutators for the transverse parts $\T_{\perp}$ and $\Tt_{\perp}$ (see eqs.(\ref{Tperp}, \ref{Tperp-flat})) can be obtained by using the results (\ref{l-ltil-comm}) and the following ones
\cite{mukhopadhyay09}: 
\bea
[l(\s), \delta \T(\s')] &=& \pi i \Pi^+(\s) \tilde \Pi^+(\s') K(\vec
X(\s')) \delta'(\D)+ {\mu\over 2} \sqrt{\pi \over T} \Pi^+(\s)
\sqrt{\tilde \Pi^+(\s')\over \Pi^+(\s')}(S\Sigma \tilde S(\s'))
\delta'(\D)~, \cr 
[\tilde l(\s), \delta \T(\s')] &=& -\pi i \Pi^+(\s') \tilde \Pi^+(\s)
K(\vec X(\s')) \delta'(\D)- {\mu\over 2} \sqrt{\pi \over T} \tilde
\Pi^+(\s) \sqrt{\Pi^+(\s')\over \tilde \Pi^+(\s')}(S\Sigma \tilde
S(\s')) \delta'(\D)~, \cr && 
\eea
into eqs.(\ref{vir-alg}). The final results are given by
eqs.(\ref{Tperp-TperpR}, \ref{Tperp-TperpL}, \ref{Tperp-TperpRL}).

\section{First order SUSY variations}
\label{a:1st}

Here we give the results for the first order susy transformations.

Here we will prove the results in eqs.(\ref{lhsR}, \ref{lhsL}, \ref{lhsRL}). The right and left moving SUSY variations of the basic fields are given by,
\bea
[\Pi^I(\s), Q_{\dot a}] &=& -i\s^I_{a\dot a} \del S^a(\s) + {i\mu \over 2\sqrt{\pi T}} (\bar \s^I\Sigma)_{\dot a a}\chi(\s) \tilde S^a(\s)~,\cr
[\tilde \Pi^I(\s), Q_{\dot a}] &=& {i\mu \over 2 \sqrt{\pi T}} (\bar \s^I\Sigma)_{\dot a a} \chi(\s) \tilde S^a(\s)~, \cr
[X^I(\s), Q_{\dot a}] &=& {i\over 2\sqrt{\pi T}} \s^I_{a\dot a} S^a(\s)~, \cr
\{S^a(\s), Q_{\dot a}\} &=& \s^I_{a\dot a} \Pi^I(\s) ~, \cr
\{\tilde S^a(\s), Q_{\dot a}\} &=& - \mu (\bar \s^I \Sigma)_{\dot a a} \chi(\s) X^I(\s)~, 
\label{basic-susy-R}
\eea
and,
\bea
[\Pi^I(\s), \tilde Q_{\dot a}] &=& -{i\mu \over 2\sqrt{\pi T}} (\bar \s^I\Sigma)_{\dot a a}\chi(\s) \tilde S^a(\s)~,\cr
[\tilde \Pi^I(\s), \tilde Q_{\dot a}] &=& i \s^I_{a\dot a} \del \tilde S^a(\s) - {i\mu \over 2\sqrt{\pi T}} (\bar \s^I\Sigma)_{\dot a a} \chi(\s) S^a(\s)~,\cr
[X^I(\s), \tilde Q_{\dot a}] &=& {i\over 2\sqrt{\pi T}} \s^I_{a\dot a} \tilde S^a(\s)~, \cr
\{S^a(\s), \tilde Q_{\dot a}\} &=& \mu (\bar \s^I\Sigma)_{\dot a a} \chi(\s) X^I(\s) ~, \cr
\{\tilde S^a(\s), \tilde Q_{\dot a}\} &=& \s^I_{a \dot a} \tilde \Pi^I(\s)~.
\label{basic-susy-L}
\eea
Using these results we compute the following SUSY variations of various components of $\T_{\perp}$ and $\Tt_{\perp}$.
\bea
[t(\s), Q_{\dot a}] &=& -i \s^I_{a\dot a} \Pi^I(\s) \del S^a(\s) + {i\mu \over 2\sqrt{\pi T}}(\bar \s^I\Sigma)_{\dot a a} \chi(\s) \Pi^I(\s) \tilde S^a(\s)~, \cr
[s(\s), Q_{\dot a}] &=& {i\over 2} \s^I_{a\dot a}\Pi^I(\s) \del S^a(\s) - {i\over 2} \s^I_{a\dot a} \del \Pi^I(\s) S^a(\s)~, \cr
[\tilde t(\s), Q_{\dot a}] &=&  {i\mu \over 2\sqrt{\pi T}}(\bar \s^I\Sigma)_{\dot a a} \chi(\s) \tilde \Pi^I(\s) \tilde S^a(\s)~, \cr
[\tilde s(\s), Q_{\dot a}] &=& {i\mu \over 2} (\bar \s^I\Sigma)_{\dot a a} \lt[\chi(\s)X^I(\s)\del \tilde S^a(\s) - \del \lt(\chi(\s)X^I(\s) \rt) \tilde S^a(\s)\rt] ~, \cr
[\tilde t(\s), \tilde Q_{\dot a}] &=& i \s^I_{a\dot a} \tilde \Pi^I(\s) \del \tilde S^a(\s) - {i\mu \over 2\sqrt{\pi T}}(\bar \s^I\Sigma)_{\dot a a} \chi(\s) \tilde \Pi^I(\s) S^a(\s)~, \cr
[\tilde s(\s), \tilde Q_{\dot a}] &=& {i\over 2} \s^I_{a\dot a}\del  \tilde \Pi^I(\s) \tilde S^a(\s) - {i\over 2} \s^I_{a\dot a} \tilde \Pi^I(\s) \del \tilde S^a(\s)~, \cr
[\delta \T_B(\s), Q_{\dot a}] &=& {i\mu^2 \over 2\sqrt{\pi T}} \s^I_{a\dot a}\chi^2(\s) X^I(\s) S^a(\s)~, \cr
[\delta \T_F(\s), Q_{\dot a}] &=& -{i\mu \over 2\sqrt{\pi T}} (\bar \s^I\Sigma)_{\dot a a} \chi(\s) \Pi^I(\s) \tilde S^a(\s) - {i\mu^2\over 2\sqrt{\pi T}} \s^I_{a\dot a} \chi^2(\s) X^I(\s) S^a(\s)~, \cr
[\delta \T_B(\s), \tilde Q_{\dot a}] &=& {i\mu^2 \over 2\sqrt{\pi T}} \s^I_{a\dot a}\chi^2(\s) X^I(\s) \tilde S^a(\s) ~, \cr
[\delta \T_F(\s), \tilde Q_{\dot a}] &=& {i\mu \over 2\sqrt{\pi T}} (\bar \s^I\Sigma)_{\dot a a} \chi(\s) \tilde \Pi^I(\s) S^a(\s) - {i\mu^2\over 2\sqrt{\pi T}}\s^I_{a\dot a} \chi^2(\s) X^I(\s) \tilde S^a(\s) ~. \cr &&
\eea

\section{Short distance behavior corresponding to MNO and PNO}
\label{a:ope}

Given the definitions of MNO and PNO in \cite{mukhopadhyay09} one can compute the short distance behavior of 
the products of the basic fields. We first take the product of two local fields at $\s$ and $\s'$ and then reorder them according to MNO and PNO. The difference can be written in terms of the functions $d(e^{\pm i\D},\eps)$, where $\D=\s-\s'$ and their derivatives. Finally, we use the following short distance behavior:
\bea
d(e^{i\D}, \eps) = {1\over 2}+{1\over \eps-i\D} + {\cal O}(\eps-i\D)~,
\eea
to find the leading terms. The non-trivial results are given below:
\bea
\Pi^I(\s)\Pi^J(\s') &=& |\Pi^I(\s)\Pi^J(\s')| + \dt^{IJ} {1\over (\eps-i\D)^2}~, 
 \label{Pi-Pi} \\ && \cr
\Pi^I(\s) X^J(\s') &=& |\Pi^I(\s) X^J(\s')| -{i\over 2\sqrt{\pi T}}\dt^{IJ} \lt({1\over 2}+{1\over \eps -i\D} \rt)  ~,
\label{Pi-X} \\ && \cr
X^I(\s) \Pi^J(\s') &=& |X^I(\s) \Pi^J(\s')|  +{i\over 2\sqrt{\pi T}} \dt^{IJ} \lt(-{1\over 2}+{1\over \eps -i\D} \rt) ~,
\label{X-Pi} \\ && \cr
\tilde \Pi^I(\s)\tilde \Pi^J(\s') &=& :\tilde \Pi^I(\s)\tilde \Pi^J(\s'): + \dt^{IJ} {1\over (\eps+i\D)^2}~, \cr
&=& \ant \tilde \Pi^I(\s)\tilde \Pi^J(\s') \ant -\dt^{IJ} {1\over (\eps-i\D)^2}~, \\ \label{Pit-Pit}
\tilde \Pi^I(\s) X^J(\s') &=& :\tilde \Pi^I(\s) X^J(\s'): -{i\over 2\sqrt{\pi T}} \dt^{IJ} \lt({1\over 2} +{1\over \eps+i\D} \rt)  ~,\cr
&=& \ant \tilde \Pi^I(\s) X^J(\s')\ant -{i\over 2\sqrt{\pi T}} \dt^{IJ} \lt({1\over 2} +{1\over \eps-i\D} \rt)~, \label{Pit-X} \\ && \cr
X^I(\s) \tilde \Pi^J(\s') &=& :X^I(\s) \tilde \Pi^J(\s'):  +{i\over 2\sqrt{\pi T}} \dt^{IJ} \lt(-{1\over 2}+{1\over \eps+i\D} \rt) ~,\cr
&=& \ant X^I(\s) \tilde \Pi^J(\s')\ant +{i\over 2\sqrt{\pi T}} \dt^{IJ} \lt(-{1\over 2}+{1\over \eps-i\D} \rt)~, \label{X-Pit} \\ && \cr
S^a(\s) S^b(\s') &=& |S^a(\s) S^b(\s')| + \dt^{ab}\lt(-{1\over 2} +{1\over \eps- i \D} \rt)~, 
\label{S-S} \\ && \cr
\tilde S^a(\s) \tilde S^b(\s') &=& :\tilde S^a(\s) \tilde S^b(\s'): + \dt^{ab}\lt(-{1\over 2} +{1\over \eps+ i \D} \rt)~, \cr
&=& \ant \tilde S^a(\s) \tilde S^b(\s')\ant +\dt^{ab} \lt(-{1\over 2} +{1\over \eps- i \D} \rt)~.\label{St-St}
\eea

\end{document}